\documentclass[11pt,twoside]{article}
\usepackage{./asp2014}

\aspSuppressVolSlug
\resetcounters

\begin{document}

\title{Probing the Sun with ALMA: observations and simulations}
\author{Maria Loukitcheva$^{1,2}$, Sami K. Solanki$^{1,3}$, Stephen M. White$^4$ and Mats Carlsson$^5$
\affil{$^1$Max-Planck-Institut for Sonnensystemforschung, G\"ottingen, Germany; \email{lukicheva@mps.mpg.de}}
\affil{$^2$Astronomical Institute, St.Petersburg University, St.Petersburg, Russia; \email{marija@peterlink.ru}}
\affil{$^3$School of Space Research, Kyung Hee University, Yongin, Gyeonggi 446-701, Korea}
\affil{$^4$Space Vehicles Directorate, Air Force Research Laboratory, Kirtland AFB, NM, United States}
\affil{$^5$Institute of Theoretical Astrophysics, University of Oslo, Oslo, Norway}}

\paperauthor{Sample~Author1}{Author1Email@email.edu}{ORCID_Or_Blank}{Author1 Institution}{Author1 Department}{City}{State/Province}{Postal Code}{Country}
\paperauthor{Sample~Author2}{Author2Email@email.edu}{ORCID_Or_Blank}{Author2 Institution}{Author2 Department}{City}{State/Province}{Postal Code}{Country}
\paperauthor{Sample~Author3}{Author3Email@email.edu}{ORCID_Or_Blank}{Author3 Institution}{Author3 Department}{City}{State/Province}{Postal Code}{Country}

\begin{abstract}
ALMA will open a new chapter in the study of the Sun by providing a leap in spatial resolution and sensitivity compared to currently available mm wavelength observations. In preparation of ALMA, we have carried out a large number of observational tests and state-of-the-art radiation MHD simulations. Here we review the best available observations of the Sun at millimeter wavelengths.
Using state of the art radiation MHD simulations of the solar atmosphere 
we demonstrate the huge potential of ALMA observations for uncovering the nature of the solar chromosphere. We show that ALMA will not only provide a reliable probe of the thermal structure and dynamics of the chromosphere, it will also open up a powerful new diagnostic of magnetic field at chromospheric heights, a fundamentally important, but so far poorly known parameter.
\end{abstract}

\section{Solar millimeter interferometric observations}

Unique solar mm observations obtained with the 10 antennas of the Berkeley-Illinois-Maryland Array (BIMA) in 2003-2004 and with the 15 antennas of the Combined Array for Research in Millimeter-wave Astronomy (CARMA) in 2009-2013 with resolution up to 10\arcsec\ have proved that interferometric observations of the solar chromosphere are feasible \citep{2006A&A...456..697W, 2006A&A...456..713L}. We demonstrate that the maximum entropy (MEM) deconvolution is a successful restoration technique for imaging of low-contrast solar targets and for oscillation power reconstruction from snapshot images. The critical issue is sampling all the spatial scales of the emission, including the largest ones. Since interferometers sample a limited range of spatial scales, one has to sacrifice spatial
resolution to achieve successful imaging. The effective spatial resolution depends on the array configuration, the observed wavelength and the position on the sky.
  
\section{Simulating ALMA observations of the chromosphere}

We used 3D radiation MHD simulations \citep{2011A&A...531A.154G} of a bipolar region of enhanced network at 0.064\arcsec\ resolution (Carlsson et al. in prep) to study the potential of chromospheric observations with ALMA \citep{2015A&A...575A..15L}.\\
We find brightness temperature (Fig.~\ref{fig1}) to be a reasonable measure of the gas temperature at the effective formation height of mm radiation (up to 1\arcsec\ resolution). The formation height range depends on the location in the simulation domain and is related to the underlying magnetic structure. We conclude that multi-wavelength ALMA observations with a small step in wavelength will provide sufficient information for tomographic imaging of the chromosphere.\\ 
We can obtain information on the magnetic field over a substantial range of chromospheric heights observing at the ALMA frequencies in dual circular polarization mode. The longitudinal magnetic field can be successfully restored from the observed degree of circular polarization and brightness temperature spectra \citep{1980SoPh...67...29B}. We find that the method reproduces the longitudinal magnetic field well with the ideal model data. The simulated circular polarization of the quiet Sun falls within $0.5$\% for the ALMA frequencies (Loukitcheva et al. in prep). Further analysis, involving realistic noise, is needed to fully validate the approach and test the requirements for the polarization observation and calibration accuracy.\\
Summarizing, ALMA will provide a unique opportunity to probe the thermal and magnetic structure of the solar atmosphere from the height of the classical temperature minimum through the chromosphere to just below the transition region.

\articlefigure[width=.85\textwidth]{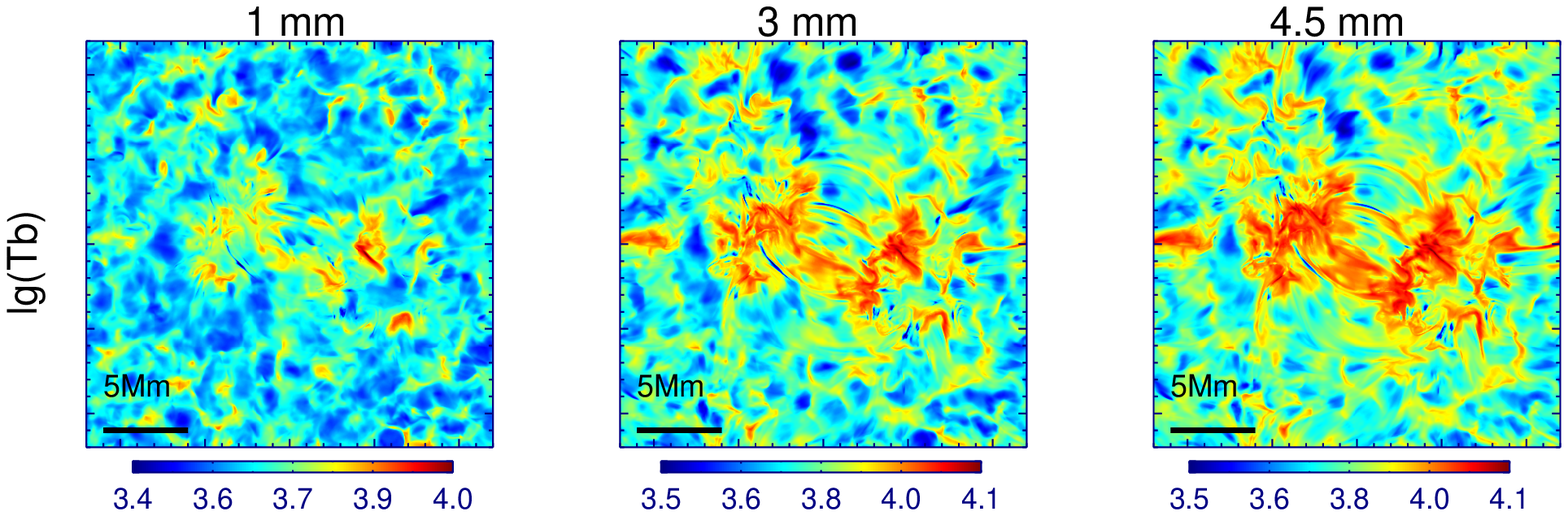}{fig1}{Synthetic mm brightness with a complex pattern of intermittent dark and bright regions, featuring bright elongated fibrils along the magnetic field lines.}

\acknowledgements This work was supported in part by RFBR grant 15-02-03835, and Saint-Petersburg State University research grants 6.0.26.2010 and 6.37.343.2015.

\bibliography{loukitcheva}  





\end{document}